%% file: VTC_Pos.tex
\documentclass[conference]{IEEEtran}
\IEEEoverridecommandlockouts
\usepackage[english]{babel}
\usepackage{amssymb,amsmath,amscd,amsfonts,amsthm,bbm,bm}
\usepackage{amsmath}
\usepackage{graphicx}
\usepackage{psfrag}
\usepackage{paralist}
\usepackage{color}
\usepackage{float}
\usepackage{relsize}
\usepackage{euscript}
\usepackage{setspace}
\usepackage{flushend}
\usepackage{tikz}
\usetikzlibrary{plotmarks}
\usepackage{pgfplots}
\usetikzlibrary{arrows,shapes,decorations,backgrounds}
\usepackage{calc}
\pgfplotsset {compat = 1.16}
\usepackage[nolist,printonlyused]{acronym}      

\def\BibTeX{{\rm B\kern-.05em{\sc i\kern-.025em b}\kern-.08em
    T\kern-.1667em\lower.7ex\hbox{E}\kern-.125emX}}

\newcommand{\bs}{\boldsymbol}

\newcommand{\R}{\mathbb R}

\newcommand{\C}{\mathbb{C}}

\newcommand{\br}{\}}

\def\bA{{\bf A}}
\def\bB{{\bf B}}
\def\bI{{\bf I}}
\def\bF{{\bf F}}

\def\bY{{\bf Y}}
\def\bH{{\bf H}}

\def\bM{{\bf M}}
\def\bV{{\bf V}}

\def\bF{{\bf F}}
\def\ba{{\bf a}}
\def\bb{{\bf b}}
\def\bc{{\bf c}}
\def\bC{{\bf C}}

\def\be{{\bf e}}

\def\bh{{\bf h}}
\def\bw{{\bf w}}

\def\bI{{\bf I}}

\def\bv{{\bf v}}
\def\bu{{\bf u}}

\def\br{{\bf r}}
\def\bs{{\bf s}}

\def\bz{{\bf z}}

\begin{document}

\title{Vehicular Positioning and Tracking \\ in Multipath Non-Line-of-Sight Channels\\
}
%

\author{
	Zhicheng Ye$^{\dag}$,
	Julia Vinogradova$^{\star}$,
	G\'{a}bor Fodor$^{\star \star \ddag}$,
    Peter Hammarberg$^{\star \star}$, \\
		
	\small $^\dag$Aalto University, Finland, Email: \texttt{zhicheng.ye@aalto.com}\\	
	\small $^{\star}$Ericsson Research, Finland, Email: \texttt{Julia.Vinogradova@ericsson.com}\\	
	\small $^{\star \star}$Ericsson Research, Sweden, E-mail: \texttt{firstname.secondname@ericsson.com} \\
	\small $^\ddag$KTH Royal Institute of Technology, Sweden. E-mail: \texttt{gaborf@kth.se}\\

}
\maketitle
\input{acronyms.tex}

\begin{abstract}
We consider the downlink transmission in a single cell multiple-input multiple-output system, in which the user equipment correspond to a vehicle moving along a given trajectory.
This system utilizes millimeter wave channels characterized by multiple non-line-of-sight (NLoS) components.
As it has been pointed out in several related works, in such systems radio access network (RAN)-based positioning can effectively improve the positioning accuracy achieved by Global Navigation Satellite Systems. However, the RAN-based positioning accuracy is highly dependent on the quality of the channel estimates, especially if multipath propagation is exploited.
Recognizing that the communication channels between the serving base station and the vehicle as well as the geographical position of the vehicle can be advantageously modeled as inter-related autoregressive processes, we propose a two-stage Kalman filter algorithm employing two intertwined filters for channel tracking, position tracking and abrupt channel change detection.
The first Kalman filter tracks angles-of-departure and angles-of-arrival associated with the communication channels, which are used to make a coarse position estimation. The second Kalman filter tracks the position of the vehicle utilizing the kinematic parameters of the vehicle. Simulation results clearly show the advantages of using the proposed scheme, which exploits the memoryful property of both the communication channels and the geographical positions, as compared to employing previously proposed single-stage or not properly combined filters in NLoS environments.
\end{abstract}

\begin{IEEEkeywords}
Kalman filter, MmWave channels, MIMO, multipath, NLoS, positioning
\end{IEEEkeywords}

\section{Introduction}
The \ac{GNSS} is one of the most widely used positioning technologies in both the civilian 
and military fields including automotive scenarios. The performance of \ac{GNSS}-based positioning techniques degrades in tunnels, 
urban canyons and other areas, in which \ac{GNSS} coverage is either poor or is not available. 
Recent advances in cellular-based positioning technology indicate that positioning techniques 
using radio access based on multiple-antenna measurements can complement GNSS-based positioning 
in vehicular scenarios in such problematic areas. 
Specifically, \ac{mmWave} signals with large antenna arrays have a potential of high accuracy positioning 
in \ac{5G} systems \cite{Wymeersch'2017}. 
This is due to large available frequency bands for a more accurate time of arrival or time-difference of arrival 
estimates that can be advantageously used to estimate range and position. 
Moreover, deploying a greater number of antennas, which is typical for \ac{mmWave} base stations, allows 
to estimate \acp{AoD} and \acp{AoA} with a higher accuracy. 
This aspect is crucial for position estimation in \ac{NLoS} environment 
in which the channel is characterized by a few dominant components.

Modern vehicles rely on a large number of sensors allowing to obtain/assist positioning. 
Sensor fusion positioning techniques combine measurements from different sources, 
such as onboard sensor and cellular measurements. 
The \acp{IMU} are widely used in vehicles and allow to determine the speed, the acceleration, 
and the direction of the vehicle for position tracking. 
For instance, a recently proposed sensor-fusion based method in \cite{Mostafavi'20} 
combines cellular \ac{mmWave} measurements with \ac{IMU} measurements. 
A Kalman filter based approach is performed at the vehicle’s side to generate its position based on the \ac{LoS} measurements.

It is important to recognize, that
positioning accuracy in \ac{mmWave} channels is highly dependent on the quality of the channel estimate. 
Indeed, the \ac{mmWave} channels are characterized by a few multipath components, not necessarily including a \ac{LoS} component. 
Therefore, an accurate channel estimation is a crucial part of designing a vehicular positioning framework. 
In the context of \ac{mmWave} channels, the channel is usually assumed to follow the $L$-scatter model, 
which is defined as a function of the number of multipath components, the \ac{AoD}  and \ac{AoA}, and the corresponding pathgains, 
where the number of multipath components depends on the surrounding channel propagation environment. 
As it has been shown in \cite{Akdeniz'2014}, in urban scenarios the channels typically comprise  
up to four multipath components. 
Channel tracking method for such a channel model was proposed in \cite{Zhang:16} using Kalman filters, 
based on a dual time scale channel model. 
In that model, the number of multipath components can change abruptly, 
and the angular variations are assumed to be slow between two abrupt changes. 
This is a particularly useful channel model for vehicular scenarios for which 
the channel can change abruptly due to fast changes in the propagation environment. 
However, the angle transition model assummed in \cite{Zhang:16}, 
for which the transition matrix is equal to identity is not realistic in practical scenarios. 
For instance, the channel aging nature of the \ac{mmWave} channels 
has been recently discussed in \cite{Yuan'2020}, \cite{Kim'2021}, \cite{Truong'2013}.

In this paper, we consider a vehicular scenario in which a vehicle follows a given trajectory. 
The channel is assumed to be $L$-scattered with $L$ dominant paths following the dual time scale model proposed in \cite{Zhang:16}. 
In this case, an \ac{AR} model for the \ac{AoD} and the \ac{AoA} evolution can be assumed, which allows to capture channel aging. 
We propose a two-stage Kalman filter algorithm relying on two intertwined Kalman filters. 
The first Kalman filter, as an extension of the work in \cite{Zhang:16}, allows to track the angular channel components and to obtain a coarse position based on the estimated channel. The second Kalman filter, based on sensor fusion using the \ac{IMU} measurements, is designed for refining the position estimation of the vehicle. The rationale for using two filters is to exploit the memoryful property of both the communication channels and the geographical positions. As we will see, this basic idea helps to overcome some of the challenges that are posed by the NLoS environment. 

The system model including the channel and transmission model is introduced in Section~\ref{System_mod}.
The channel tracking is presented in ~\ref{channelest}. 
The overall algorithm is provided in Section~\ref{pos_est} 
including a coarse position estimation and refined position tracking. 
Simulations results are discussed in Section~\ref{simu}. Conclusions are drawn in Section~\ref{concl}.

\textit{Notations:} The superscripts $(\cdot)^{\sf T}$ and $(\cdot)^{\sf H}$ denote transpose and Hermitian transpose, respectively. The notation ${\cal CN}(0, \sigma^2)$ represent the complex circular Gaussian distributions with mean $a$ and variance $\sigma^2$. The vector $\boldsymbol{0}_{n}$ denotes the $n$-dimensional vector with entries equal to zero and $\bI_{n}$ denotes the identity matrix of dimension $n \times n$. The notation $\text{vec} \left\{ \cdot \right\}$ corresponds to the vectorization operator.

\section{System model}\label{System_mod}

\subsection{Channel model}

\begin{figure}[ht]
	\centering
\includegraphics[width=8cm]{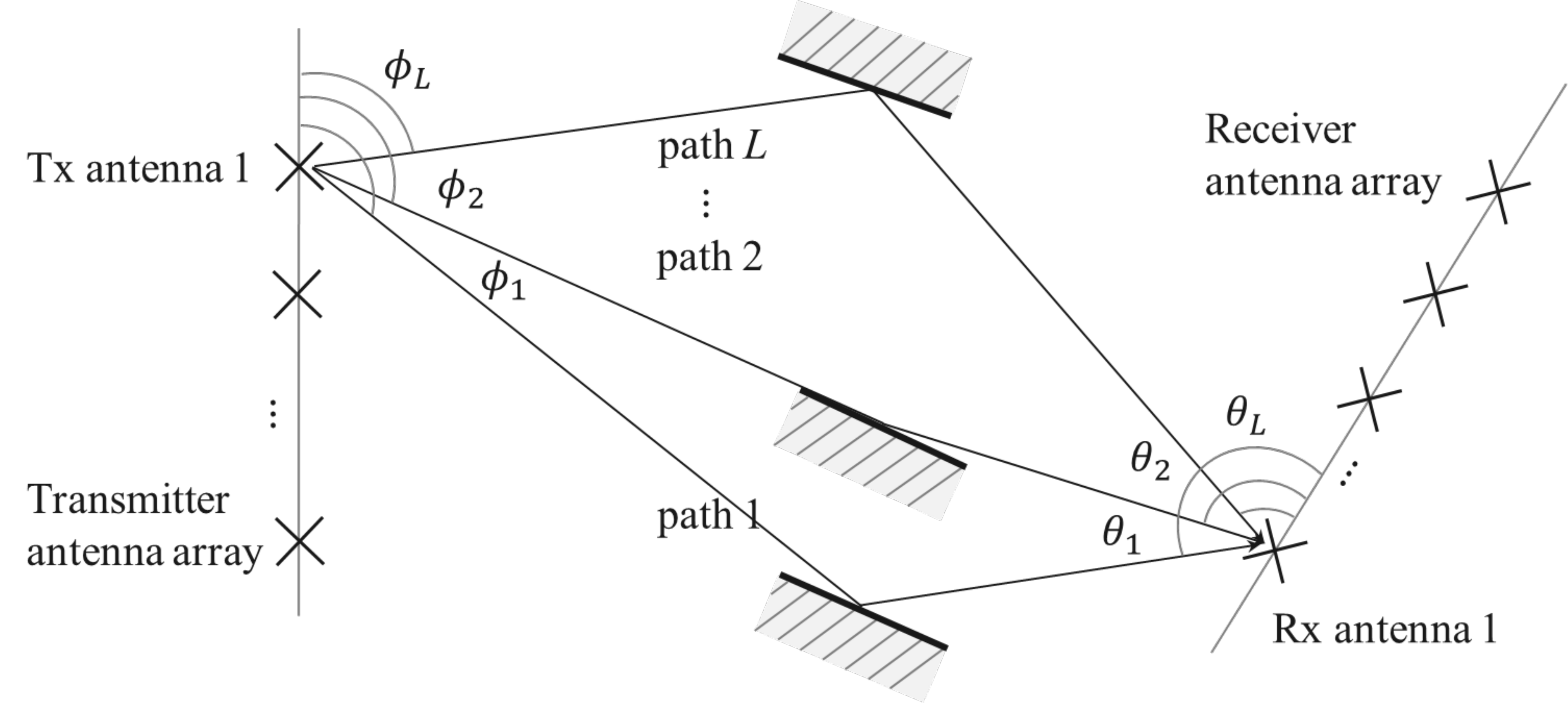}
	\caption{L-scatterer channel model.}
	\label{fig:channel}
\end{figure}
We consider a downlink system with a single \ac{BS} equipped with a linear array 
of $N_t$ transmit antennas and \ac{UE} equipped with a linear array of $N_r$ receive antennas. 
We assume the $L$-scatterer channel model \cite{Zhang:16}, as depicted in Figure~\ref{fig:channel}, 
where $\phi_l$ and $\theta_l$ are the \ac{AoD}  and the \ac{AoA}, respectively, 
for $l=1, \ldots, L$, where $L$ is the number of multipath components. 
The corresponding steering vectors are defined by
\begin{align}
	\textbf{a}_t(\phi)&\triangleq \frac{1}{\sqrt{N_t}} \left[1, e^{-j\pi \cos \phi}, \ldots, e^{-j\pi (N_t-1)\cos \phi} \right]^{\sf T} \label{steer1} \\
	\textbf{a}_r(\theta)&\triangleq \frac{1}{\sqrt{N_r}} \left[1, e^{-j\pi \cos \theta}, \ldots, e^{-j\pi (N_r-1)\cos \theta} \right]^{\sf T} \label{steer2}.
\end{align}
The $L$-scatterer channel is then suitably defined as
\begin{equation}\label{H}
	\bH=\sum_{l=1}^L \alpha_l \textbf{a}_r(\theta_l)\textbf{a}_t^{\sf H}(\phi_l),
\end{equation}
where $\ba_t(\phi) \in \C^{N_t \times 1}$ and $\ba_r(\theta) \in \C^{N_r \times 1}$ 
are defined in \eqref{steer1} and \eqref{steer2}, 
$\alpha_l=\rho_l\sqrt{N_t N_r} e^{-j \frac{2\pi \Delta_l}{\lambda_c}}$ 
with $\lambda_c$ the wavelength, and $\alpha_l$, $\rho_l$, $\Delta_l$ are 
the path gain, the attenuation, and the distance between transmit antenna 1 
and receive antenna 1 along the path $l$, respectively.

We define the following vector that combines all the \ac{AoD}s and the \ac{AoA}s as
\begin{equation}\label{psi}
	\boldsymbol{\psi}=[\phi_1, \ldots, \phi_L, \theta_1, \ldots, \theta_L]^{\sf T} \in \C^{2L \times 1}.
\end{equation}
We present a method for estimating $\boldsymbol{\psi}$ in Section~\ref{channelest}.

\subsection{Transmission model}
We assume the quantization levels of both the transmitter and the receiver precoders are equal to the number of transmit and receive antennas $N_t$ and $N_r$, respectively. 
The beam-forming and beam-combining vectors can then be expressed as:
\begin{equation*}
	\bb = \frac{1}{\sqrt{N_t}} \begin{bmatrix}e^{j\nu_1}, e^{j\nu_2}, \dots, e^{j\nu_{N_t}}\end{bmatrix}^{\sf T} \in \C^{N_{\text{t}} \times 1},
\end{equation*}
\begin{equation*}
	\bc = \frac{1}{\sqrt{N_r}} \begin{bmatrix}e^{j\mu_1}, e^{j\mu_2}, \dots, e^{j\mu_{N_r}}\end{bmatrix}^{\sf T} \C^{N_r \times 1},
\end{equation*}
where $\nu_{n_t} \in [0, 2\pi ]$ with ${n_t} = 1, \dots, N_t$ and $\mu_{n_r} \in [0, 2\pi ]$ with ${n_r} = 1, \dots, N_r$.

We further assume that the $N_t$ beamforming vectors and the $N_r$ combining vectors cover the range $[0, \pi]$.
The transmitter sends pilots using beamforming vectors 
$\textbf{b}_{n_t}=\textbf{a}_t(\bar{\phi}_{n_t})$ 
for $n_t=1, \ldots, N_t$. 
For each beamforming vector, the receiver uses the combining vectors 
$\bc_{n_r}=\textbf{a}_r(\bar{\theta}_{n_r})$ for $n_r=1, \ldots, N_r$. 
For convenience, but without losing generality, we will assume that the same pilot symbol $x$ is sent over all the $N_t$ transmitter antennas. 
Then, the same observation scalar is received over $N_r$ receiver antennas:
\begin{equation*}
	y_{n_r n_t}={\bc}_{n_r}^{\sf H} \bH {\textbf{b}}_{n_t}+{\bc}^{\sf H}_{n_r} \bw_{n_t},
\end{equation*}
where $x=1$, $\bw_{n_t} \in \C^{N_r\times 1}$ is an additive white Gaussian noise with zero mean and variance $\sigma_w^2$, and $\bH \in \C^{N_r \times N_t}$ is defined in \eqref{H}.

The $N_r \times N_t$ observation matrix is written as
\[
\bY=
\left (
\begin{array}{cccc}
	y_{11} & y_{12}& \cdots &y_{1N_t} \\
	y_{21} & y_{22}& \cdots &y_{2N_t} \\
	\vdots & \vdots& \ddots & \vdots \\
	y_{N_r1} & y_{N_r2}& \cdots &y_{N_rN_t} \\
\end{array}
\right ) = \textbf{C}^{\sf H} \bH \textbf{B}+\bV,
\]
where $\bC=[\bc_{1}, \bc_{2}, \ldots, \bc_{N_r}] \in \C^{N_r \times N_r}$ and $\bB=[\textbf{b}_{t,1}, \textbf{b}_{t,2}, \ldots, \textbf{b}_{t,N_t}] \in \C^{N_t \times N_t}$, and $\bV=[\bv_{1}, \ldots, \bv_{N_t}] \in \C^{N_r \times N_t}$ with $\bv_{n_t}= \bc^{\sf H}_{n_r} \bw \sim {\cal CN}(\boldsymbol{0}_{N_r}, \sigma^2_w \bI_{N_r})$ for $n_t=1, \ldots, N_t$.

\section{Channel tracking using Kalman filter}\label{channelest}

In the Kalman filter framework, we define the state vector by $\boldsymbol \psi$ as in \eqref{psi}.
We assume that the $\boldsymbol \psi (t)$ is a complex Gaussian stationary process 
following the autoregressive model of order $p\geq1$ in time, denoted by AR($p$).

The state transition equation at time $t$ is defined by
\begin{equation}\label{psi_tran}
	\boldsymbol{\psi}(t)= \sum_{i=1}^{p}\bA_i\boldsymbol{\psi}(t-p) + \textbf{u}(t),
\end{equation}
where $\bu(t) \sim {\cal CN}(\boldsymbol{0}_{2L}, \sigma_u^2\bI_{2L}) \in \C^{2L \times 1}$ is the process noise
vector and the matrices 
$\bA_i \in \C^{2L \times 2L}$ for $i \in \left\{1, \ldots, p\right\}$ 
are the state transition matrices assumed to be constant in time.

The measurement equation at time instant $t$ is given by
\begin{equation}\label{yn}
	\textbf{y}(t)=\widetilde{\textbf{h}}\left(\boldsymbol{\psi}(t)\right) + \widetilde{\bv}(t),
\end{equation}
where $\textbf{y}(t) = \text{vec} \left\{ \bY \right\} \in \C^{N_rN_t \times 1}$, $\widetilde{\bh}\left(\boldsymbol{\psi}\right)= \text{vec} \left\{ \textbf{C}^{\sf H} \bH \textbf{B} \right\}\in \C^{N_rN_t \times 1}$ and 
$\widetilde{\bv}(t)= \text{vec} \left\{ \bV \right\}\sim {\cal CN}(\boldsymbol{0}_{N_rN_t}, \sigma^2_w \bI_{N_rN_t})\in \C^{N_rN_t \times 1}$ is the  measurement noise vector.

In the Kalman filter framework, the term 
$\widetilde{\textbf{h}}\left(\boldsymbol{\psi}\right)$ 
is a nonlinear function of the state vector $\boldsymbol{\psi}$, 
and hence an extended Kalman filter framework is applied, similarly as in \cite{Zhang:16}, and the explicit details are omitted here.
The \textit{a posteriori} estimate of the channel state vector 
at time instant $t$ given $t$ observations is denoted by $\widehat{\boldsymbol{\psi}}(t|t)$.

\section{Position estimation and tracking}\label{pos_est}

\subsection{Position estimation in multipath channels with single point scatterers} \label{coarse}
We consider a two-dimensional space with a single \ac{BS} located at the point $(x_{\text{BS}}, y_{\text{BS}})$ and the UE located at $(x_{\text{UE}}, y_{\text{UE}})$ in $(x,y)$-coordinate system with the orientation denoted by $\gamma$ relative to the \ac{BS}.
We consider $\theta_l$, $\phi_l$, and $R_l$ to be the \ac{AoA}, the \ac{AoD}, the pathlength, respectively, corresponding to the single-point scatterer for the $l$th multipath component for $l=1, \ldots, L$. We propose a method to estimate $(x_{\text{UE}}, y_{\text{UE}}, \gamma)$, given $\theta_l$, $\phi_l$, and $R_l$ for $l=1, \ldots, L$.

\begin{figure}[ht] 
	\centering
\includegraphics[width=0.75\hsize]{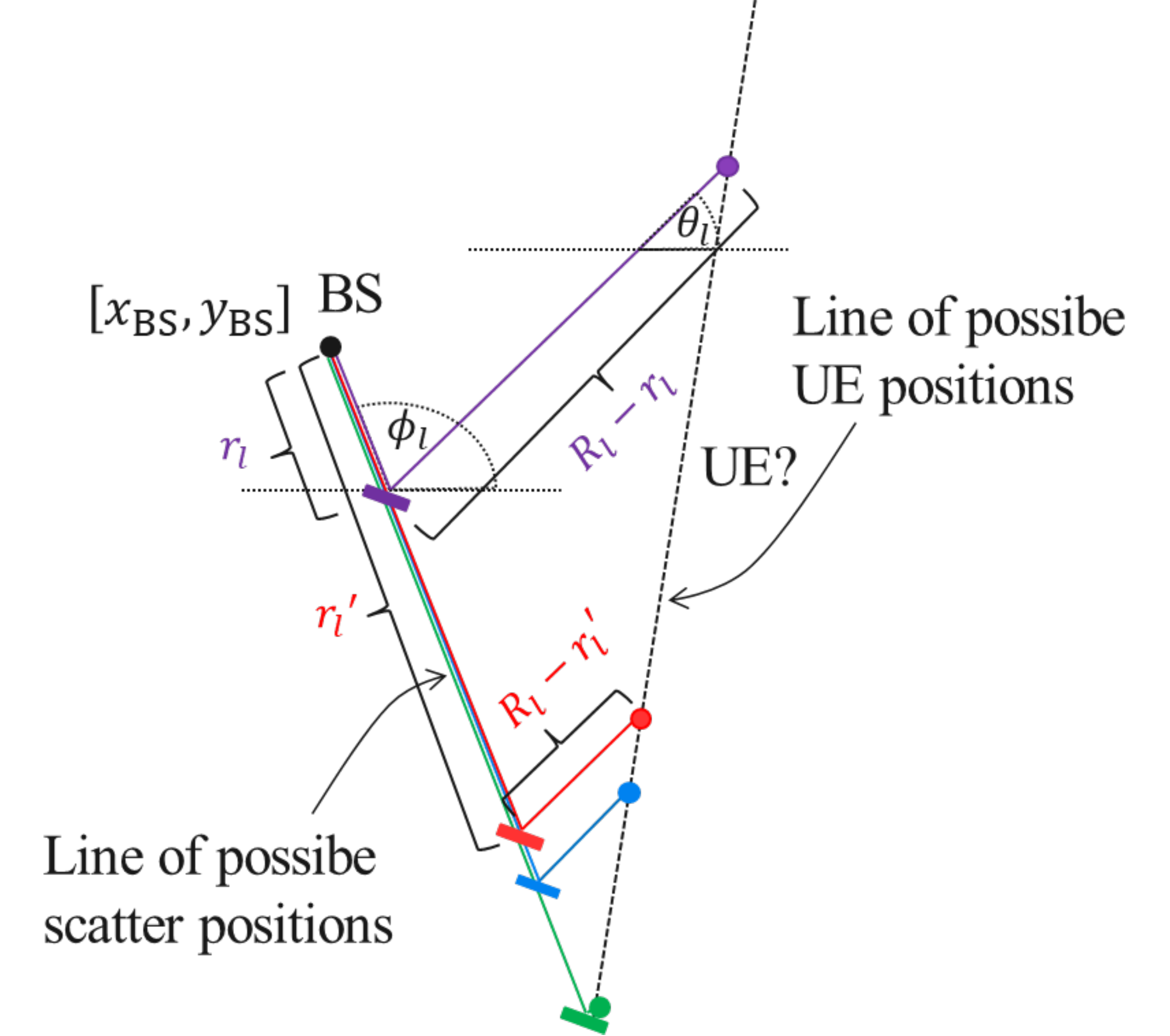}
	\caption{The triangulation positioning diagram. Each \ac{NLoS} path defines one line segment on which the \ac{UE} is expected to be located.}
	\label{fig:triangulation}
\end{figure}

The \ac{UE} is located on the line defined by the following system of equations, as illustrated in Figure~\ref{fig:triangulation}, for any $0<r_l<R_l$
\begin{align*}
\begin{cases}
x(r_l) &= r_l \text{cos}\phi_l \text{cos} (\theta_l +\gamma) +(R_l-r_l) +x_{\text{BS}}	\\
y(r_l) &= r_l \text{sin}\phi_l \text{sin} (\theta_l +\gamma) +(R_l-r_l) +y_{\text{BS}}.
\end{cases}
\end{align*}
Equivalently, for each $l$th scatterer with $l=1, \ldots, L$, the equation of the \ac{UE} location line is defined by
\begin{equation*}
	0=a_lx+b_ly+c_l
\end{equation*}
	with $a_l$, $b_l$, and $c_l$ the solutions of the system of equations:
\begin{align*}
	\begin{cases}
		a_l &= \text{sin}\phi_l- \text{sin}(\theta_l +\gamma)	\\
		b_l &= \text{cos} (\theta_l +\gamma) - \text{cos}\phi_l  \\
		c_l &= -a_l\left(R_l\text{cos} (\theta_l +\gamma)+x_{\text{BS}}\right)-b\left(R_l\text{sin} (\theta_l +\gamma)+y_{\text{BS}}\right).
	\end{cases}
\end{align*}
Note that if all the above parameters are known perfectly, the location of the \ac{UE} is given by the intersection between the $L$ lines, which is the true position $(x,y)$. In practice, due to the imperfect estimates of the \ac{AoA}s, \ac{AoD}s, and pathlengths, these lines do not intersect in a single point.
We define the following cost function:
\begin{equation*}
	f(x,y,\gamma)=\sum_{l=1}^L \beta_l d_l^2,
\end{equation*}
where $\beta_l$ is the weight taking into account 
any non-equal reliability of the different measurement sets and $d_l$ is the distance between the point $(x,y)$ and the $l$th line defined by
\begin{equation*}
	d_l=\frac{|a_lx+b_ly+c_l|}{\sqrt{a_l^2+b_l^2}}.
\end{equation*}

The estimate of the \ac{UE} position/orientation is obtained by minimizing the cost function $f(x, y, \gamma)$:
	\begin{equation}\label{opt_pb}
(x^*, y^*, \gamma^*) ={\text{arg min}}_{x,y,\gamma}	f(x, y, \gamma).
	\end{equation}

The minimum to $f(x,y,\gamma)$ is found by setting the gradient with respect to $x$, $y$ and $\gamma$ to zero.

\subsection{Two-stage Kalman filter for position tracking} \label{posKF}

Once a one-shot position and orientation estimation has been obtained by using Equation~\eqref{opt_pb}, a more refined position can be obtained by exploiting tracking by means of a Kalman filter. We define the \ac{UE} position state vector at time $t$ as
\begin{equation*}
	\bs(t) = [x(t), y(t), \dot{x}(t), \dot{y}(t), \ddot{x}(t), \ddot{y}(t), \gamma(t)]^{\sf T} \in \R^{7\times1},
\end{equation*}
where $x(t)$, $y(t)$, $\dot{x}(t)$, $\dot{y}(t)$, $\ddot{x}(t)$, $\ddot{y}(t)$ are respectively the \ac{UE}'s positions, the velocities, and the accelerations in $x$ and $y$ coordinates, and $\gamma(t)$ is the orientation of the \ac{UE} at time instant $t$.

The state transition and the measurement equations are given by
\begin{align}\label{posKFeq}
		\bs(t) &= \bF \bs(t-1) + \be(t),	\nonumber \\
		\bz(t) &= \bM \bs(t) + \br(t),
\end{align}
where $\bF$ is the $7 \times 7$ state transition matrix of a constant acceleration process, $\bM=\bI_{7}$ is the measurement matrix, $\be(t) \sim {\cal N}(\boldsymbol{0}_7, \sigma_{e}^2 \bI_{7})$ is the process noise vector, and $\br(t) \sim {\cal N}(\boldsymbol{0}_7, \sigma_{r}^2 \bI_{7})$ is the measurement noise vector.

\subsection{Two-stage Kalman filter for position tracking}

\begin{figure}[h!]
	\begin{center}
		\includegraphics[width=6.5cm]{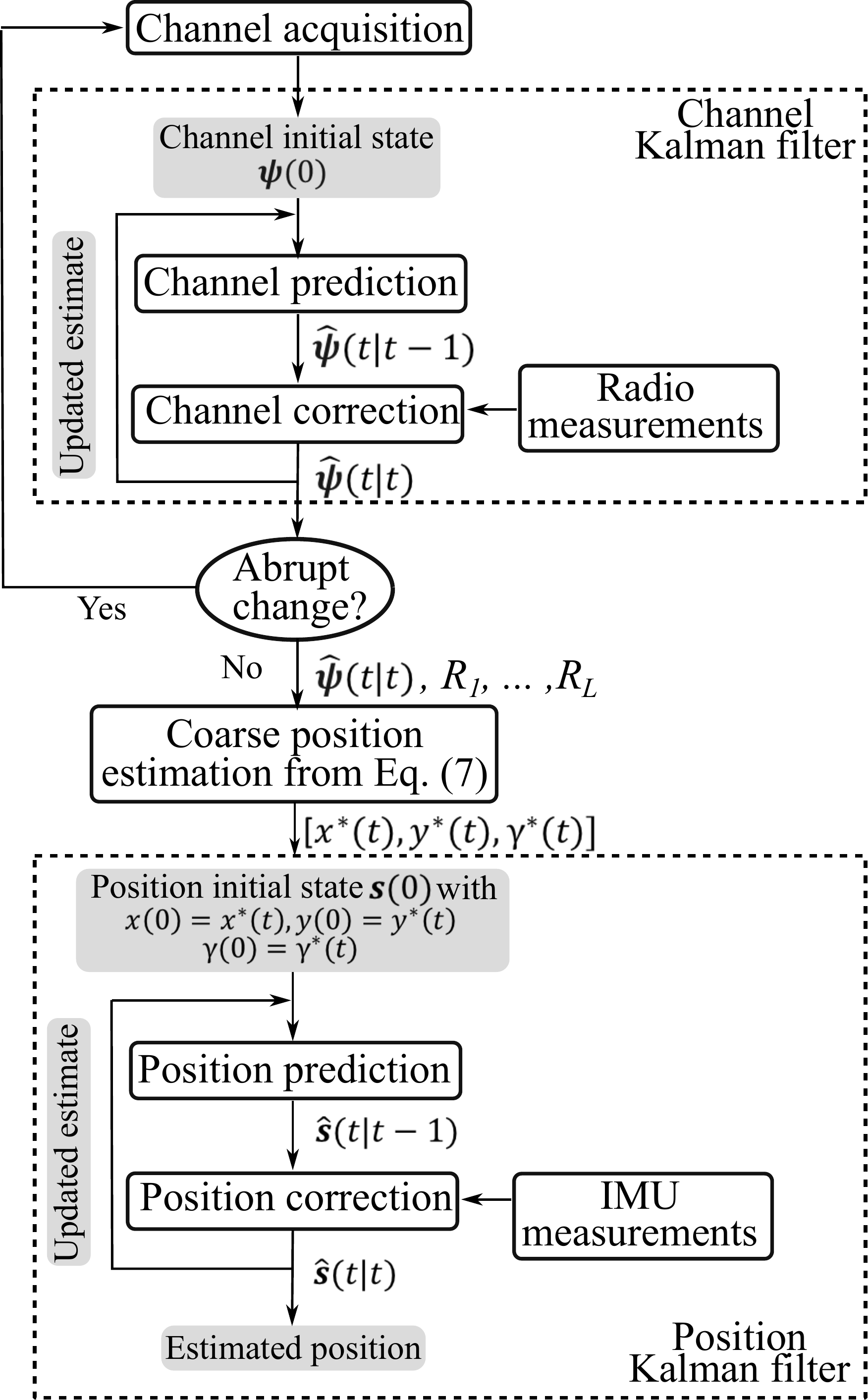}
		\caption{Flow-chart for position estimation using two-stage Kalman filter.}
		\label{FlowChart}
	\end{center}
\end{figure}

Multipath channel estimation is a crucial component in \ac{mmWave} radio systems that needs to be used in a large number of system modules in addition to the positioning module. Therefore, we propose a two-stage Kalman filter algorithm in order to track the \ac{UE}'s position independently from the channel tracking as depicted in Figure~\ref{FlowChart}.

In dense urban environments it is natural to assume that the number of multipath components may vary abruptly as compared to the variations of the \ac{AoD}/\ac{AoA} and the pathgains whose variations are considered to be slow. This is due, for instance, to a sudden change of a scatterer's location or a sudden blockage of a path. We consider the double time scale channel variation model, similar to the one proposed in \cite{Zhang:16}, for which the number of multipath components $L$ remains constant between two abrupt changes. As depicted in Figure~\ref{FlowChart}, the first Kalman filter is used to estimate the channel state vector assuming $L$ constant if no abrupt changes is detected and the estimate, obtained at time instant $t$ given $t$ observations, is denoted by $\widehat{\boldsymbol{\psi}}(t|t)$. At each step, an abrupt change detection test \cite{Zhang:16} based on a function of the measurement equation is performed given a certain probability of false alarm. If an abrupt change is detected, the new number of multipath components $L$ is considered by performing channel acquisition for which $L$ is re-estimated. The prediction and correction steps are performed using the transition and measurement equations defined in \eqref{psi_tran} and \eqref{yn}, respectively. Once the channel is estimated, the estimate $\widehat{\boldsymbol{\psi}}(t|t)$ along with the path lengths $R_l$ (assumed to be known at each time istant $t$), for $l=1, \ldots, L$, are used to estimated a coarse position as described in Section~\ref{coarse}. A second parallel Kalman filter is used in order to track the \ac{UE}'s position such as described in Section~\ref{posKF}. For the second Kalman filter, the prediction and measurement equations are used such as defined in Equation~\eqref{posKFeq}.

\section{Simulation results}\label{simu}
\begin{figure}[h] 
	\centering
	\includegraphics[width=8cm]{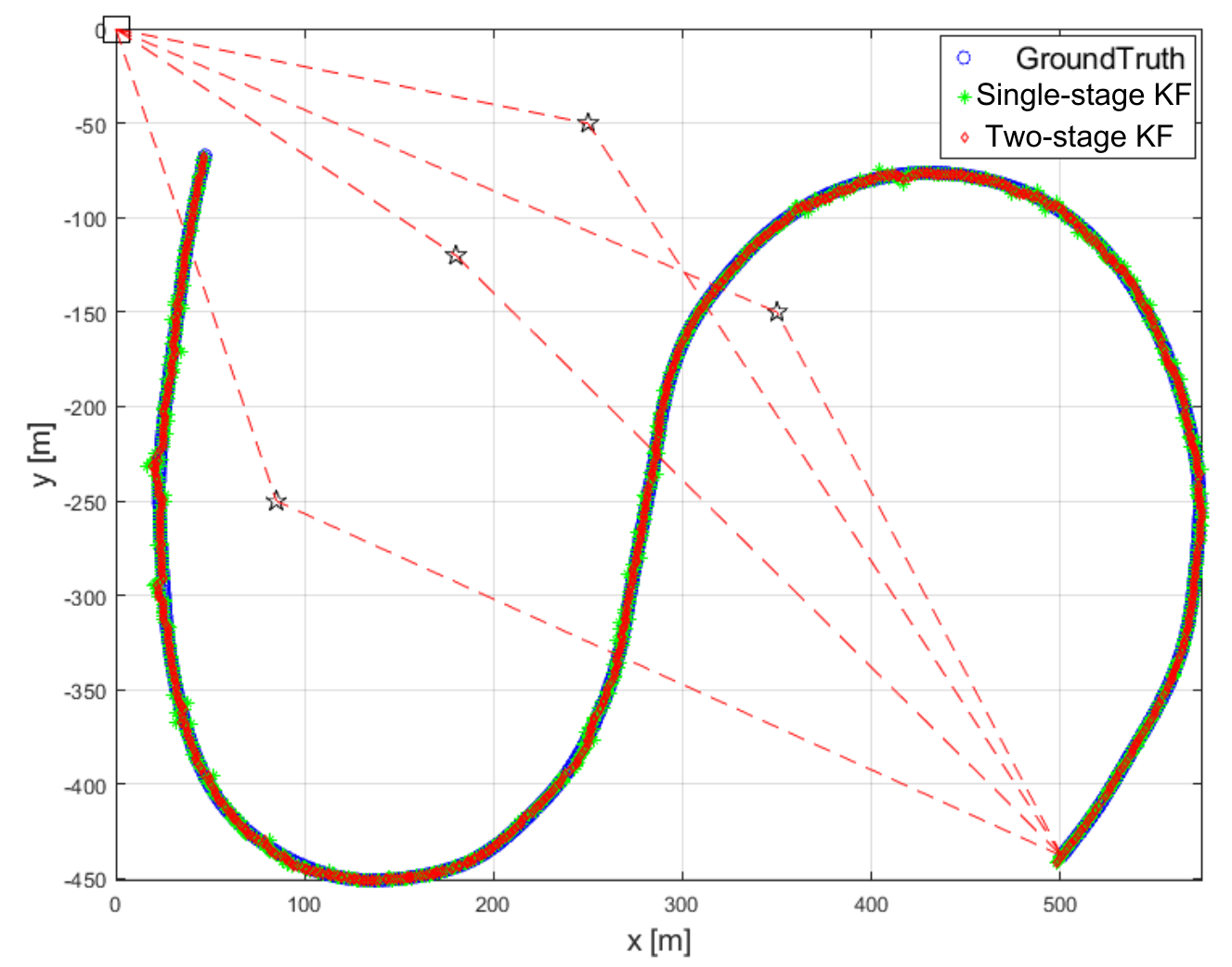}
		\vspace{-0.4cm}
	\caption{Vehicle's trajectory and the estimated positions.}
	\label{fig:trajectory}
\end{figure}

We consider a suburban scenario with a moving vehicle representing the \ac{UE}.
The simulations have been carried out in MATLAB using the Driving Scenario Designer toolbox 
allowing a customized driving trajectory. 
The S-shape trajectory, as depicted Figure~\ref{fig:trajectory}, has been used. 
The trajectory data including the positions, the orientation, 
and the velocity of each sampling time has been saved and then later used 
for the proposed algorithm performance analysis. 
As depicted in Figure~\ref{fig:trajectory}, the \ac{BS} is placed at $x_{\text{BS}}=0, y_{\text{BS}}=0$ 
in the upper left corner of the 500-by-600-meter rectangular area where the vehicle is allowed to move. 
The relative distance between the vehicle and the \ac{BS} ranges from 0 to 800 meters approximately.

A \ac{MIMO} system is considered with linear antenna arrays at the \ac{BS} 
with 64 antennas and the \ac{UE} with 8 antennas. 
The BS is operating at the carrier frequency equal to 40 GHz. 
A downlink with $L=4$ multipath components is assumed motivated by the experimental results provided in~\cite{Akdeniz'2014}.
The \ac{AoD}/\ac{AoA} are assumed to follow the same \ac{AR} process of order 1 with coefficient $a_1=0.95$ 
corresponding to $\bA_1=a_1 \bI_{2L}$ in \eqref{psi_tran}. 
The \ac{SNR} is assumed to be equal to 20 dB, $\sigma_u^2=(0.5\pi/180)^2$ and $\sigma_w^2=N_rN_t/\text{SNR}$ as in \cite{Zhang:16}.
The scatterers are randomly placed around the UE every 50 meters. At each new channel acquisition, we assume the initial channel state vector to be equal to the true channel corrupted by a Gaussian error with variance $\sigma_u^2=(0.5\pi/180)^2$. The path lengths $R_1, \ldots, R_L$ are assumed to be known and correspond to the genie range measurements.
The vehicle is assumed to be moving with a constant velocity of $54$ km/h.

\begin{figure}[h] 
	\centering
	\includegraphics[width=8cm]{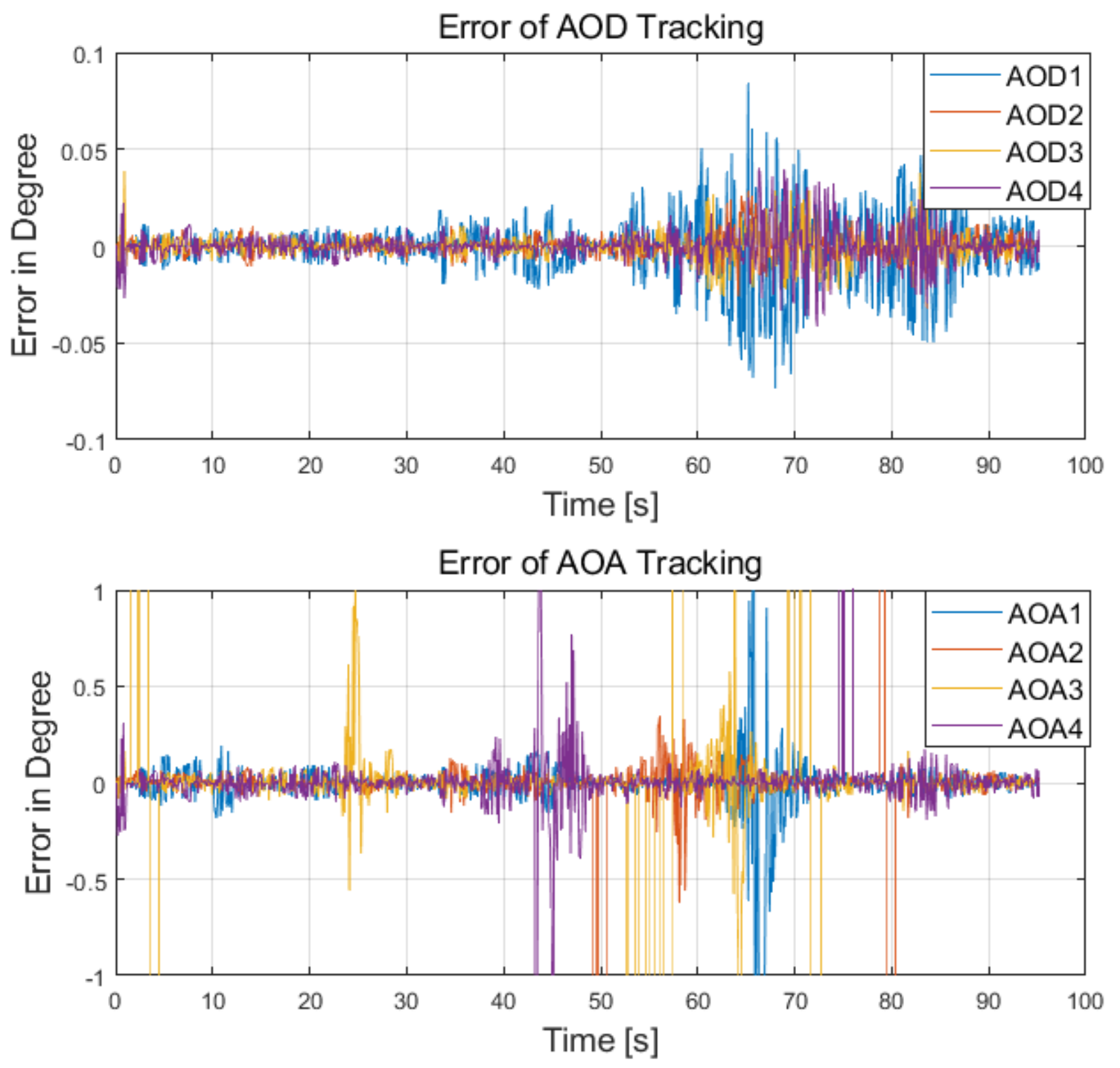}
		\vspace{-0.4cm}
	\caption{\ac{AoD}/\ac{AoA} mean square estimation errors.}
	\label{fig:angles}
\end{figure}

We compare the proposed two-stage Kalman filter-based method denoted 
by 'Two-stage KF' with the method based on the coarse position estimation using only the first channel tracking Kalman filter, denoted by 'Single-stage KF'. The corresponding trajectory and the scatterer's positions 
at the final time instant are depicted in Figure~\ref{fig:trajectory}. 
The channel tracking in terms of \ac{AoD}/\ac{AoA} estimation errors are provided in Figure~\ref{fig:angles}. 
We note that larger errors are obtained for the \acp{AoA} as compared to the \acp{AoD}. 
This is due to the fact that given the single-point scatterer model, the \acp{AoD} 
do not change when the scatterers' positions are constant.

\begin{figure}[h] 
	\centering
	\includegraphics[width=8cm]{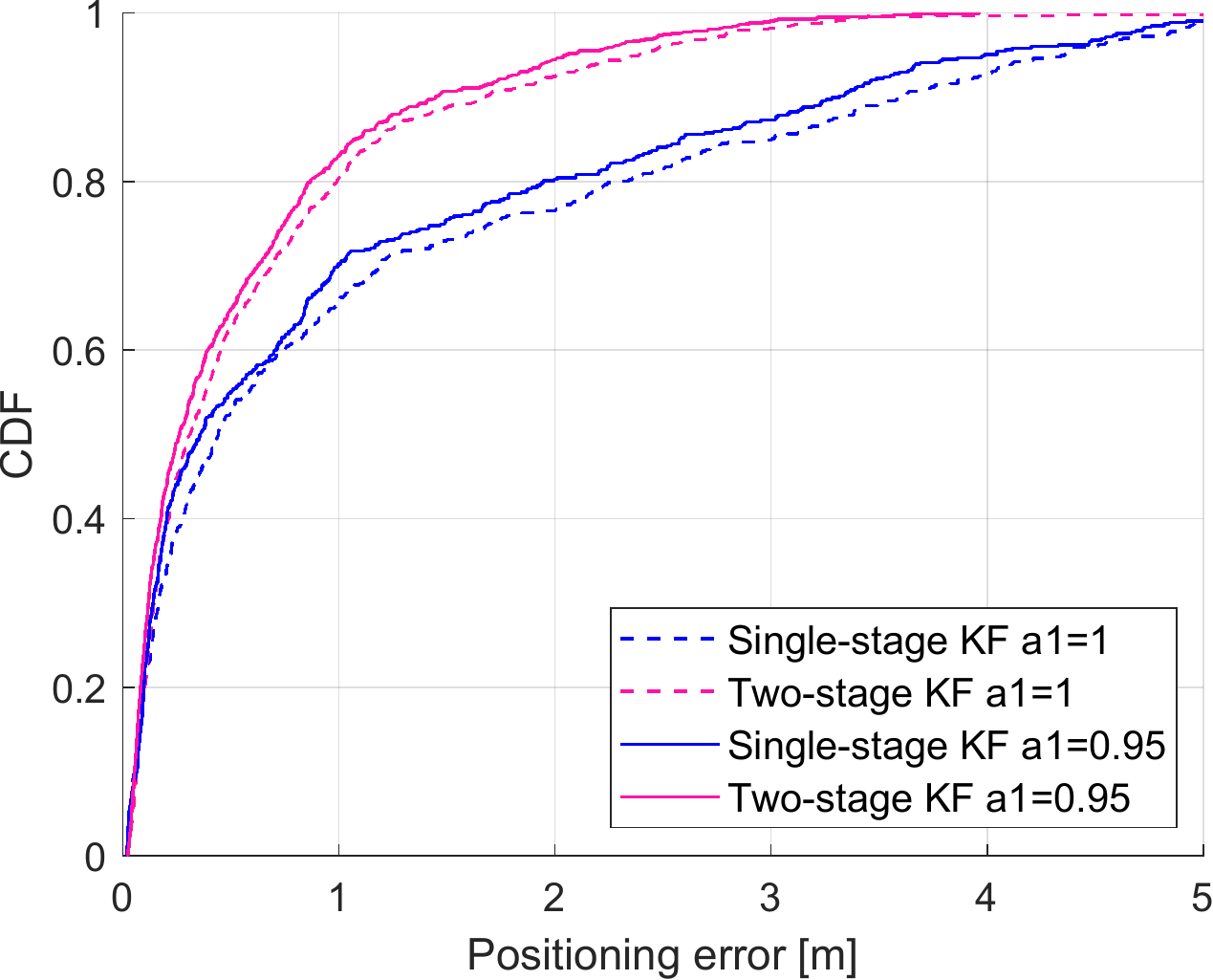}
	\vspace{-0.2cm}
	\caption{\ac{CDF} of the positioning error for the single-stage and two-stage methods.}
	\label{fig:CDF}
\end{figure}

In Figure~\ref{fig:CDF}, the \ac{CDF} curves for the single- and two-stage Kalman filter are compared for two channel state transitions models with coefficients $a_1=1$ and $a_1=0.95$. We note that the model with $a_1=1$ corresponds to the channel model considered in \cite{Zhang:16}. We observe that the results show a better performance of the two-stage approach as compared to the single-stage method. Moreover, a gain in performance of more than a decimeter is obtained by considering $a_1=0.95$ as compared to the channel estimation approach in \cite{Zhang:16} for which $a_1=1$. This shows that the choice of the transition matrix has a significant impact on positioning performance, and further studies are needed to better understand the behavior.

\section{Conclusions}\label{concl}
This paper proposed a two-stage Kalman filter that employs two intertwined
filters for channel tracking, position tracking and abrupt channel state detection. 
The rationale of this scheme is that both the vehicle's geometric position and
its communication channel can be advantageously modeled as autoregressive processes,
whose respective states can be tracked and predicted by Kalman filters.  
Specifically, the first Kalman filter tracks angles-of-departure
and angles-of-arrival associated with the communication channels
and helps to make a coarse position estimation, while the second Kalman filter 
tracks the position of the vehicle.
Numerical results indicate the advantages of using the
proposed scheme, compared to employing previously proposed single-stage or not
properly combined filters in NLoS environments. Moreover, the channel state transition matrix choice can largely affect the positioning performance and should be further investigated.

\bibliographystyle{IEEEtran}
\bibliography{Bibliography}

\end{document}

%% file: acronyms.tex

\begin{acronym}[LTE-Advanced]
  \acro{2G}{Second Generation}
  \acro{3G}{3$^\text{rd}$~Generation}
  \acro{3GPP}{3$\text{rd}$~Generation Partnership Project}
  \acro{4G}{4$^\text{th}$~Generation}
  \acro{5G}{5$^\text{th}$~Generation}
  \acro{5GPPP}{5G Infrastructure Public Private Partnership}
  \acro{QAM}{quadrature amplitude modulation}
  \acro{ADAS}{Advanced driver assistance system}
  \acro{AD}{autonomous driving}
  \acro{AI}{artificial intelligence}
  \acro{AoA}{angle-of-arrival}
  \acro{AoD}{angle-of-departure}
  \acro{API}{application programming interface}
  \acro{AR}{autoregressive}
  \acro{ARQ}{automatic repeat request}
  \acro{BER}{bit error rate}
  \acro{BLER}{block error rate}
  \acro{BPC}{Binary Power Control}
  \acro{BPSK}{Binary Phase-Shift Keying}
  \acro{BRA}{Balanced Random Allocation}
  \acro{BS}{base station}
  \acro{CAM}{cooperative awareness messages}
  \acro{CAP}{Combinatorial Allocation Problem}
  \acro{CAPEX}{capital expenditure}
  \acro{CBF}{coordinated beamforming}
  \acro{CBR}{congestion busy ratio}
  \acro{CDD}{cyclic delay diversity}
  \acro{CDF}{cumulative distribution function}
  \acro{CDL}{clustered delay line}
  \acro{CS}{Coordinated Scheduling}
  \acro{C-ITS}{cooperative intelligent transportation system}
  \acro{CSI}{channel state information}
  \acro{CSIT}{channel state information at the transmitter}
  \acro{D2D}{device-to-device}
  \acro{DCA}{Dynamic Channel Allocation}
  \acro{DCI}{downlink control information}
  \acro{DE}{Differential Evolution}
  \acro{DENM}{decentralized environmental notification messages}
  \acro{DFO}{Doppler frequency offset}
  \acro{DFT}{Discrete Fourier Transform}
  \acro{DIST}{Distance}
  \acro{DL}{downlink}
  \acro{DMA}{Double Moving Average}
  \acro{DMRS}{Demodulation Reference Signal}
  \acro{D2DM}{D2D Mode}
  \acro{DMS}{D2D Mode Selection}
  \acro{DMRS}{demodulation reference symbol}
  \acro{DPC}{Dirty paper coding}
  \acro{DPS}{Dynamic point switching}
  \acro{DRA}{Dynamic resource assignment}
  \acro{DSA}{Dynamic spectrum access}
  \acro{eMBB}{enhanced mobile broadband}
  \acro{eV2X}{Enhanced vehicle-to-everything}
  \acro{EIRP}{equivalent isotropically radiated power}
  \acro{ERTMS}{European Rail Traffic Management System}
  \acro{ETSI}{European Telecommunications Standards Institute}
  \acro{FDD}{frequency division duplexing}
  \acro{FR1}{frequency range-1}
  \acro{FR2}{frequency range-2}
  \acro{GNSS}{global navigation satellite system}
  \acro{HARQ}{hybrid automatic repeat request}
  \acro{HST}{high-speed train}
  \acro{IAB}{integrated access and backhaul}
  \acro{ITS}{intelligent transportation system}
  \acro{KPI}{key performance indicator}
  \acro{IEEE}{Institute of Electronics and Electrical Engineers}
  \acro{IMT}{International Mobile Telecommunications}
  \acro{IMU}{inertial measurement unit}
  \acro{InC}{in-coverage}
  \acro{IoT}{Internet of Things}
  \acro{ITS}{intelligent transportation system}
  \acro{LDPC}{low-density parity-check coding}
  \acro{LMR}{land mobile radio}
  \acro{LoS}{line-of-sight}
  \acro{LTE}{Long Term Evolution}
  \acro{MAC}{medium access control}
  \acro{mmWave}{millimeter-wave}
  \acro{MBB}{mobile broadband}
  \acro{MCS}{modulation and coding scheme}
  \acro{METIS}{Mobile Enablers for the Twenty-Twenty Information Society}
  \acro{MIMO}{multiple-input multiple-output}
  \acro{MISO}{multiple-input single-output}
  \acro{ML}{machine learning}
  \acro{MRC}{maximum ratio combining}
  \acro{MS}{mode selection}
  \acro{MSE}{mean square error}
  \acro{MTC}{machine type communications}
  \acro{multi-TRP}{multiple transmission and reception points}
  \acro{mMTC}{massive machine type communications}
  \acro{cMTC}{critical machine type communications}
  \acro{NDAF}{Network Data Analytics Function}
  \acro{NF}{network function}
  \acro{NR}{New Radio}
  \acro{NLoS}{non-line-of-sight}
  \acro{NSPS}{national security and public safety}
  \acro{NWC}{network coding}
  \acro{OEM}{original equipment manufacturer}
  \acro{OFDM}{orthogonal frequency division multiplexing}
  \acro{OoC}{out-of-coverage}
  \acro{PSBCH}{physical sidelink broadcast channel}
  \acro{PSFCH}{physical sidelink feedback channel}
  \acro{PSCCH}{physical sidelink control channel}
  \acro{PSSCH}{physical sidelink shared channel}
  \acro{PDCCH}{physical downlink control channel}
  \acro{PDCP}{packet data convergence protocol}
  \acro{PHY}{physical}
  \acro{PLNC}{physical layer network coding}
  \acro{PPPP}{proximity services per packet priority}
  \acro{PPPR}{proximity services per packet reliability}
  \acro{PSD}{power spectral density}
  \acro{RLC}{radio link control}
  \acro{QAM}{quadrature amplitude modulation}
  \acro{QCL}{quasi co-location}
  \acro{QoS}{quality of service}
  \acro{QPSK}{quadrature phase shift keying}
  \acro{PaC}{partial coverage}
  \acro{RAISES}{Reallocation-based Assignment for Improved Spectral Efficiency and Satisfaction}
  \acro{RAN}{radio access network}
  \acro{RA}{Resource Allocation}
  \acro{RAT}{Radio Access Technology}
  \acro{RB}{resource block}
  \acro{RF}{radio frequency}
  \acro{RS}{reference signal}
  \acro{RSRP}{Reference Signal Received Power}
  \acro{SA}{scheduling assignment}
  \acro{SFN}{Single frequency network}
  \acro{SNR}{signal-to-noise ratio}
  \acro{SINR}{signal-to-interference-plus-noise ratio}
  \acro{SC-FDM}{single carrier frequency division modulation}
  \acro{SFBC}{space-frequency block coding}
  \acro{SCI}{sidelink control information}
  \acro{SL}{sidelink}
  \acro{SLAM}{simultaneous localization and mapping}
	\acro{SPS}{semi-persistent scheduling}
  \acro{STC}{space-time coding}
  \acro{SW}{software}
  \acro{TCI}{transmission configuration indication}
  \acro{TBS}{transmission block size}
  \acro{TDD}{time division duplexing}
  \acro{TRP}{transmission and reception point}
  \acro{TTI}{transmission time interval}
  \acro{UAV}{unmanned aerial vehicle}
  \acro{UAM}{urban air mobility}
  \acro{UE}{user equipment}
  \acro{UL}{uplink}
  \acro{URLLC}{ultra-reliable and low latency communications}
  \acro{VUE}{vehicular user equipment}
  \acro{V2I}{vehicle-to-infrastructure}
  \acro{V2N}{vehicle-to-network}
  \acro{V2X}{vehicle-to-everything}
  \acro{V2V}{vehicle-to-vehicle}
  \acro{V2P}{vehicle-to-pedestrian}
  \acro{ZF}{Zero-Forcing}
  \acro{ZMCSCG}{Zero Mean Circularly Symmetric Complex Gaussian}
 \acro{TBS}{transport block size}
 \acro{SCI}{sidelink control information}
\end{acronym}

%% file: VTC_Pos.bbl
\begin{thebibliography}{1}
\providecommand{\url}[1]{#1}
\csname url@samestyle\endcsname
\providecommand{\newblock}{\relax}
\providecommand{\bibinfo}[2]{#2}
\providecommand{\BIBentrySTDinterwordspacing}{\spaceskip=0pt\relax}
\providecommand{\BIBentryALTinterwordstretchfactor}{4}
\providecommand{\BIBentryALTinterwordspacing}{\spaceskip=\fontdimen2\font plus
\BIBentryALTinterwordstretchfactor\fontdimen3\font minus
  \fontdimen4\font\relax}
\providecommand{\BIBforeignlanguage}[2]{{%
\expandafter\ifx\csname l@#1\endcsname\relax
\typeout{** WARNING: IEEEtran.bst: No hyphenation pattern has been}%
\typeout{** loaded for the language `#1'. Using the pattern for}%
\typeout{** the default language instead.}%
\else
\language=\csname l@#1\endcsname
\fi
#2}}
\providecommand{\BIBdecl}{\relax}
\BIBdecl

\bibitem{Wymeersch'2017}
H.~Wymeersch, G.~Seco-Granados, G.~Destino, D.~Dardari, and F.~Tufvesson,
  ``5{G} mmwave positioning for vehicular networks,'' \emph{IEEE Wireless
  Communications}, vol.~24, no.~6, pp. 80--86, 2017.

\bibitem{Mostafavi'20}
S.~S. Mostafavi, S.~Sorrentino, M.~B. Guldogan, and G.~Fodor, ``Vehicular
  positioning using 5{G} millimeter wave and sensor fusion in highway
  scenarios,'' in \emph{ICC 2020 - 2020 IEEE International Conference on
  Communications (ICC)}, 2020, pp. 1--7.

\bibitem{Akdeniz'2014}
M.~Akdeniz, Y.~Liu, M.~Samimi, S.~Sun, S.~Rangan, T.~Rappaport, and E.~Erkip,
  ``\BIBforeignlanguage{English (US)}{Millimeter wave channel modeling and
  cellular capacity evaluation},'' \emph{\BIBforeignlanguage{English (US)}{IEEE
  Journal on Selected Areas in Communications}}, vol.~32, no.~6, pp.
  1164--1179, Jun. 2014.

\bibitem{Zhang:16}
C.~{Zhang}, D.~{Guo}, and P.~{Fan}, ``Tracking angles of departure and arrival
  in a mobile millimeter wave channel,'' in \emph{2016 IEEE International
  Conference on Communications (ICC)}, 2016, pp. 1--6.

\bibitem{Yuan'2020}
J.~{Yuan}, H.~Q. {Ngo}, and M.~{Matthaiou}, ``Machine learning-based channel
  prediction in massive {MIMO} with channel aging,'' \emph{IEEE Transactions on
  Wireless Communications}, vol.~19, no.~5, pp. 2960--2973, 2020.

\bibitem{Kim'2021}
H.~{Kim}, S.~{Kim}, H.~{Lee}, C.~{Jang}, Y.~{Choi}, and J.~{Choi}, ``Massive
  {MIMO} channel prediction: Kalman filtering vs. machine learning,''
  \emph{IEEE Transactions on Communications}, vol.~69, no.~1, pp. 518--528,
  2021.

\bibitem{Truong'2013}
K.~T. {Truong} and R.~W. {Heath}, ``Effects of channel aging in massive {MIMO}
  systems,'' \emph{Journal of Communications and Networks}, vol.~15, no.~4, pp.
  338--351, 2013.

\end{thebibliography}
